\documentclass{IET-Conf}
\usepackage{amsmath,amsfonts}
\usepackage{algorithmic}
\usepackage{algorithm}
\usepackage{array}
\usepackage{multirow}
\usepackage{textcomp}
\usepackage{stfloats}
\usepackage{url}
\usepackage{verbatim}
\usepackage{graphicx}
\usepackage[dvipsnames]{xcolor}
\usepackage[inkscapelatex=false]{svg}
\usepackage{subcaption}
\usepackage{float}
\usepackage{orcidlink}  

\def\BibTeX{{\rm B\kern-.05em{\sc i\kern-.025em b}\kern-.08em
    T\kern-.1667em\lower.7ex\hbox{E}\kern-.125emX}}
\begin{document}

\title{INFLUENCE OF CONVERTER CURRENT LIMITING AND PRIORITIZATION ON PROTECTION OF HIGHLY IBR-PENETRATED NETWORKS}

\author{Andrés E Quintero\ad{1}\corr, Vinícius A Lacerda\ad{1}, Oriol Gomis-Bellmunt\ad{1}, Moisés J. B. B. Davi\ad{2}, Mario Oleskovicz\ad{2}}

\address{\add{1}{CITCEA-UPC, Polytechnic University of Catalonia, Barcelona, Spain}
\add{2}{University of São Paulo, Department of Electrical Engineering, São Carlos, Brazil}
\email{andres.eduardo.quintero@upc.edu}\vspace{-10 pt}}


\begin{abstract}
This paper investigates how grid-forming (GFM) and grid-following (GFL) control strategies in inverter-based resources (IBRs) influence line distance and differential protection in converter-dominated transmission systems. A modified IEEE 39-bus system is evaluated with GFM and GFL units equipped with low-voltage ride-through logic, current limiting, and positive- or negative-sequence prioritization. Distance protection is implemented with a mho characteristic, while line differential protection uses an alpha-plane approach. Results show that phase-to-ground loops in distance protection can substantially overestimate the fault location near the Zone-1 reach. For line differential protection, external faults may cause the operating point to briefly enter the trip region of the alpha-plane, even for the healthy-phase in ABG faults under GFL control and during the initial moments of the fault, demanding strong external security measures. These findings highlight that modern converter controls, together with current limitation and sequence-current prioritization, can compromise the reliability and security of traditional protection schemes.
\end{abstract}

\maketitle

\section{Introduction}

The transition towards renewable-dominated power systems has introduced a steady increase in inverter-based resources (IBRs) in transmission networks. This transition has resulted in challenges for the operation and protection of these systems \cite{10374177}, \cite{IEEE_PES_TR81_2020}. IBRs are power-electronic interfaces that exhibit different fault current behavior, thus not following the behavior expected from traditional synchronous generators. Furthermore, in actual power systems these IBRs are integrated into the networks primarily through two control strategies, grid-following (GFL) and grid-forming (GFM), with the latter introduced to overcome limitations of conventional GFL approaches \cite{9982455}. This integration has not occurred without challenges. Specifically, line protection such as distance, directional, and phase-selection algorithms are being affected \cite{SEL1}, \cite{8666728}, \cite{10542885}.

The operational differences between synchronous generators and inverter-based resources (IBRs) are essential to evaluating the performance of traditional protection schemes. A critical distinction lies in the limited short-circuit contribution of IBRs, which typically remains close to their nominal rating \cite{9973369}. To ensure converter integrity, various current-limiting strategies have been proposed and implemented—ranging from control in the stationary frame to the synchronous reference frame \cite{ZAREI2021107020}, \cite{10362015}, \cite{10842930}. These dynamics have been recently studied \cite{10807525}, \cite{10689179} and they also influence the dynamic response and system stability during disturbances and its capacity to support the system during low-voltage ride-through (LVRT) events. This has led to efforts to seek improved control strategies for GFM and GFL converters during LVRT \cite{9948153}. Consequently, this change of paradigm is directly impacting relay sensitivity, selectivity, and fault detection reliability.

To better understand how protection functions are affected by GFL and GFM control, several works have attempted to address this matter \cite{10547428}, \cite{11193857}. The recent contributions make clear the current need to understand how protection systems are affected by converters. Moreover, a crucial aspect is the existence of many approaches to control and operate GFM and GFL converters, resulting in a wider scope and greater challenge for researchers, making the analysis harder since there is no standard approach and the implementation is usually up to the developer, where only compliance with certain requirements is mandatory. However, most works tend to analyze the response of the system considering only grid-following or grid-forming control approaches, leaving room for a more comprehensive review considering both solutions.

The performance of protection functions is also affected by the capability of an IBR to inject positive- and/or negative-sequence currents during LVRT events of asymmetrical faults, which is determined by its control design. This aspect is relevant given that some IBRs are limited to positive-sequence injection only, which restricts the fault signatures detectable by protection devices and reduces the information available for accurate fault classification and localization \cite{10902358}. Moreover, implementing GFL controls that can handle only positive-sequence injection has been shown to negatively affect negative-sequence directional overcurrent and quadrilateral distance protection elements \cite{10547428}.

Even after the efforts to understand the impacts of high IBR penetration in modern power systems, there are topics that still need further exploration. For instance, how current limiting and prioritization affect different protection functions is scarcely studied, considering that the selected method will directly affect the electrical quantities available for relays to operate. Furthermore, line differential protection is usually considered to be blinded against these new scenarios and is not often studied. This work therefore proposes to explore how current limitation and prioritization strategies could affect the operation of line distance and alpha-plane differential protection.

\section{Converter control schemes: GFM and GFL} \label{control}

\begin{figure}[]
    \vspace{-10 pt}
    \centering
    \includegraphics[width=0.62\columnwidth]{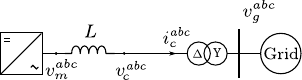}
    \caption{Grid-forming converter interface to the AC network}
    \label{fig:GFMCircuit}
\end{figure}

Given the critical influence of the control scheme on protection algorithm performance, as highlighted in recent studies, the GFM and GFL strategies, including the current-limiting and prioritization modules, were implemented to assess line protection behavior. The converter connection at the transmission level is shown in Fig.~\ref{fig:GFMCircuit}. Each unit connects to the high-voltage terminal $v_g^{abc}$ through a delta–wye transformer, synthesizes the terminal voltage $v_m^{abc}$, and measures the current $i_c^{abc}$ and voltage $v_c^{abc}$ after the filter inductance $L$.

\begin{figure}[b]
    \centering
    \includegraphics[width=0.95\columnwidth]{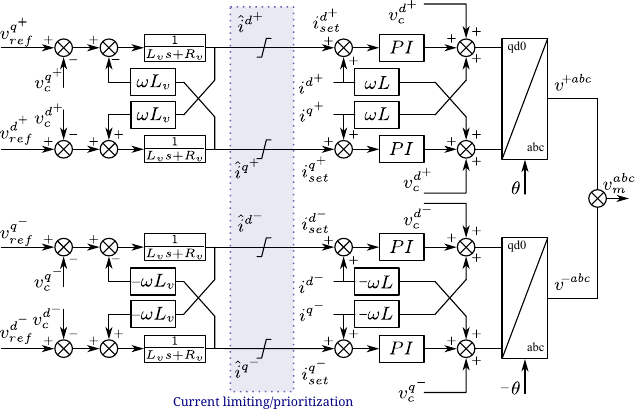}
    \caption{Control structure of the grid-forming unit}
    \label{fig:GFMVA}
\end{figure}

The GFM control strategy consists of a cascaded control formed by an outer voltage loop that generates current references, which then pass through the inner current-control loop. The control is performed in decoupled positive- and negative-sequence $qd$ frames, enabling independent control of both sequences, which is important for assessing the performance of the protection functions. The inner loop employs typical PI controls, while the outer loop uses a virtual admittance that responds to grid-voltage changes; this is shown in Fig. \ref{fig:GFMVA}.

Virtual-admittance voltage control has attracted notable interest recently \cite{9640078}, \cite{9189190}, exhibiting favorable small-signal stability margins for GFM operation against strong grids, often exceeding those of virtual-impedance schemes \cite{9662150}. The parameters $R_v$ and $L_v$ correspond to the admittance parameters, while the term $\omega L_v$ captures the cross-coupling between the $q$ and $d$ axes. Their absolute values and ratio are key determinants of closed-loop stability \cite{9662150}.

\begin{figure}[h]
    \centering

    \begin{subfigure}[t]{0.48\columnwidth}
        \centering
        \includegraphics[width=\linewidth]{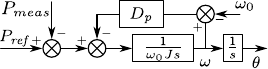}
        \caption{Virtual synchronous generator}
        \label{fig:net1}
    \end{subfigure}
    \hfill
    \begin{subfigure}[t]{0.48\columnwidth}
        \centering
        \includegraphics[width=\linewidth]{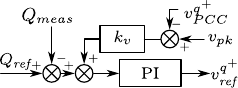}
        \caption{Reactive power droop}
        \label{fig:net2}
    \end{subfigure}

    \caption{P and Q control schemes}
    \label{fig:nets}
\end{figure}

Active- and reactive-power regulation in the GFM unit are achieved through virtual synchronous generator and reactive droop control schemes, respectively, as shown in Fig.~\ref{fig:net1} and Fig.~\ref{fig:net2}. The active-power loop provides inertial behavior and smooth convergence to the reference power, while the reactive-power loop maintains voltage regulation at the PCC through a proportional–integral controller.

Within this architecture, the virtual-admittance stage delivers the unconstrained references $\hat{i}^{qd^{\pm}}$, after which LVRT requirements and a current-limiting block produce the final references $i^{qd^{\pm}}$ delivered to the inner current controllers. 

\begin{figure}[b!]
    \centering
    \includegraphics[width=0.95\columnwidth]{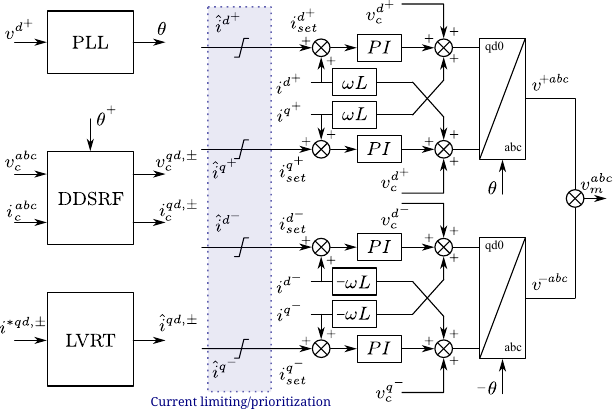}
    \caption{Control structure of the grid-following unit}
    \label{fig:VA}
\end{figure}

The grid-following control, shown in Fig. \ref{fig:VA}, is simpler and includes an inner current loop and a PLL locked to the $v_d$ component, with voltage and current $qd$ components obtained via a decoupled double synchronous reference frame (DDSRF). The reference current $i^{*qd^{\pm}}$ is modified by the LVRT logic, complying with reactive-current requirements, and then passed to the current-limiting and prioritization stage.

\section{Case Study: Highly IBR-penetrated IEEE 39-bus system}

\begin{figure}[t]
    \centering
    \includegraphics[width=0.85\columnwidth]{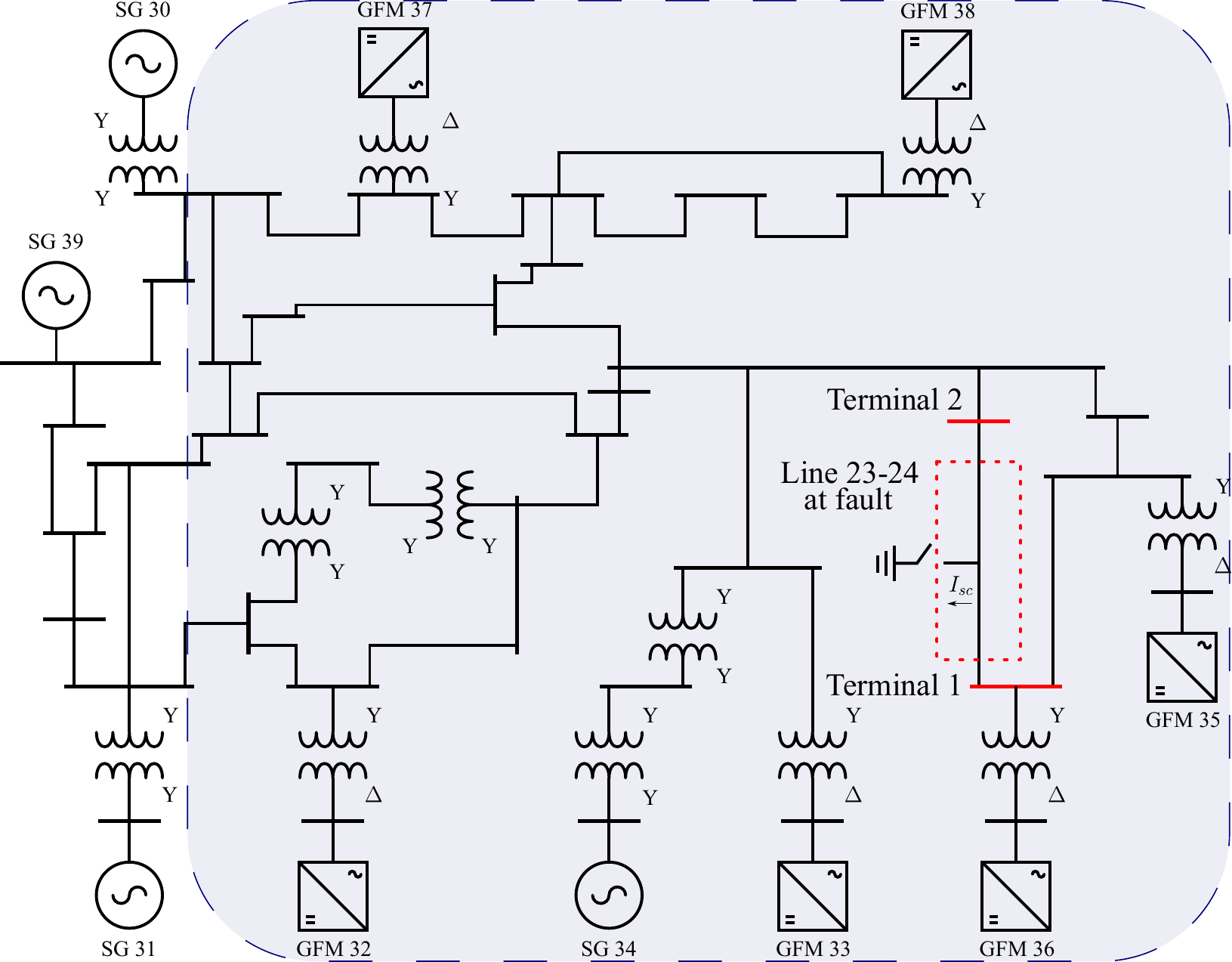}
    \caption{IEEE 39-bus with embedded GFM/GFL units}
    \label{fig:39bus}
    \vspace{-10 pt}
\end{figure}

The IEEE 39-bus benchmark in Fig. \ref{fig:39bus} was reconfigured for high penetration of inverter-based resources and simulated using electromagnetic transient (EMT) models in Matlab/Simulink. The reference case consists of ten synchronous generators; however, because more than half of the generation is achieved through converters, the short-circuit contribution and, more critically, the overall dynamic response during disturbances become predominantly shaped by the converters’ control rather than by machine electromechanics.

In the modified network, six grid-forming (GFM) units replace part of the original synchronous generation, resulting in an area highly penetrated by converters.

\section{Methodology}

For assessing the operation of the distance and differential line protection, several scenarios will be studied by faulting the line 23–24 shown in Fig. \ref{fig:39bus}. The highly IBR-penetrated zone is highlighted in blue, where the majority of the short-circuit current to the fault will come from IBRs. Regarding the protection functions, the distance protection employs the well-known mho characteristic, and the differential algorithm is based on the alpha-plane, using $\alpha = I_L/I_R$ to map the operating point in the complex plane. The line differential algorithm is known for its high reliability. To deliver a complete analysis, the algorithm will be tested against different fault scenarios for line-to-ground faults and double-line-to-ground faults, where the unbalanced fault naturally induces negative-sequence components, as well as different fault resistances and fault locations, including internal and external faults.

As previously stated, the implemented strategy for the control system of the converters will play a crucial role in the response of the protection functions. To address this, the system will be analyzed for scenarios with GFM and GFL control strategies. Moreover, current prioritization is also considered, testing for positive- or negative-sequence priority during LVRT events. Finally, the current limitation algorithm ensures that the currents remain within the converter-rated current limits.

\section{Results}

This section briefly evaluates a few scenarios resulting from the performance of the protection algorithms, according to the system shown in Fig. \ref{fig:39bus} and the control system of Section \ref{control}.

\subsection{Distance protection}

\begin{figure}[t]
    \centering
    \includegraphics[width=1\columnwidth]{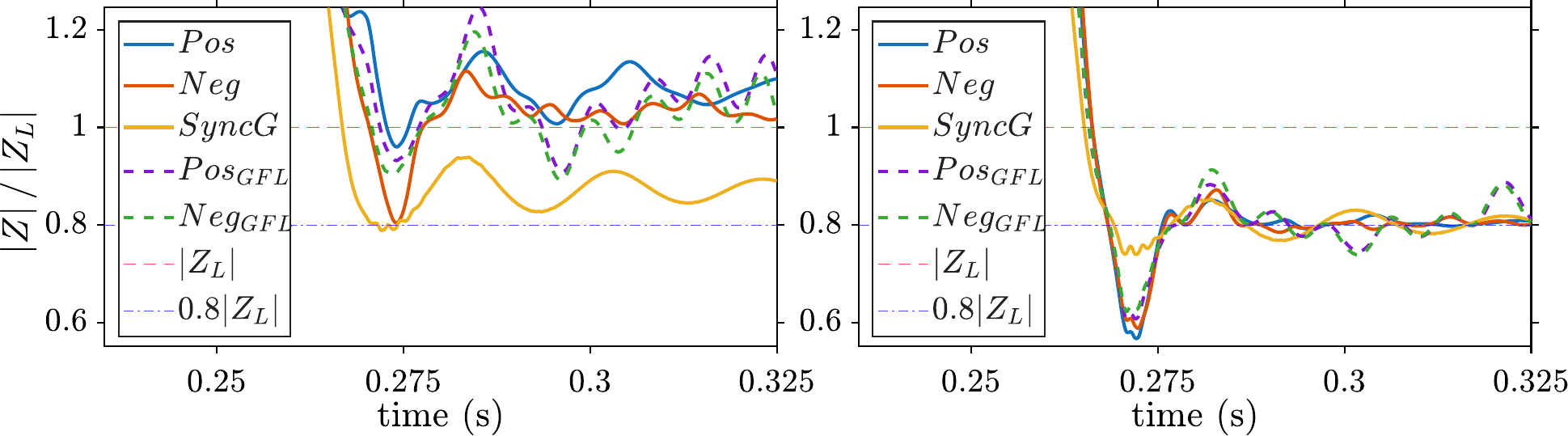}
    \caption{Left: AG loop. Right: AB loop. ABG fault at 80\%}
    \label{fig:dist80}
\end{figure}

The performance of the distance protection was evaluated by analyzing the calculated apparent impedance under different fault scenarios. Fig.~\ref{fig:dist80} illustrates an ABG fault located at 80\% of the line length with a fault resistance of $Z_F \approx 0$. The left plot presents the trajectory for the AG loop, while the right plot shows that for the AB loop. In both cases, the impedance is compared against the relay reach (dashed blue line).

\begin{figure}[b]
    \centering
    \includegraphics[width=1\columnwidth]{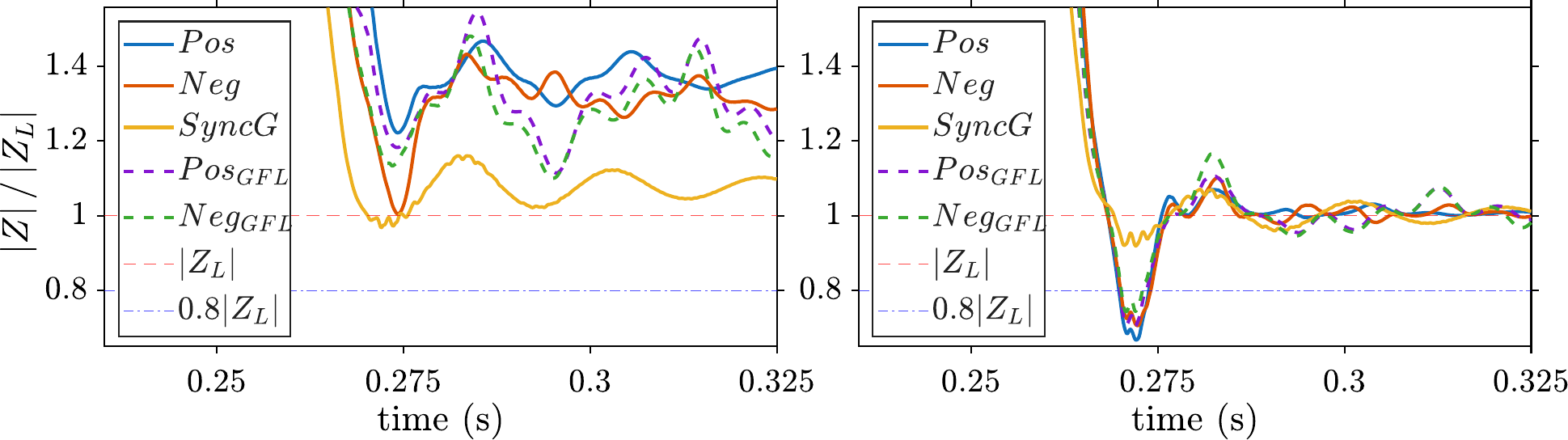}
    \caption{Left: AG loop. Right: AB loop. ABG fault at 100\%}
    \label{fig:dist100}
\end{figure}
The calculated impedance obtained from the purely synchronous generation scenario (solid yellow line) is the closest to the actual apparent impedance. In contrast, for the GFM and GFL cases, whether prioritizing positive- or negative-sequence currents, the apparent impedance reaches significantly higher values  ranging from 81\% to 125\% of the line length for the AG loops. This deviation is most noticeable in the AG loops, where the zero-sequence current influences the loop impedance calculation. Even when applying the traditional compensation factor $k_0 = \frac{Z_{0L} - Z_{1L}}{3Z_{1L}}$, the use of a single fixed $k_0$ value cannot prevent the apparent impedance from exhibiting such large variations, with an overall range of nearly 50\% of the line length.

When comparing positive- and negative-sequence prioritization, it is clear that with positive prioritization the apparent impedance is larger than when prioritizing negative sequence. This difference is expected to be even larger for converters with only positive-sequence injection, which tend to elevate the positive- and negative-sequence voltage during LVRT events. Following the same idea, when analyzing the GFL and GFM control behavior, the value of the impedance tends to oscillate more aggressively when the system is supported by GFL units.

It can be concluded that for AG loops, the calculated impedance tends to reach higher values than the actual fault location when using typical $k_0$ values, whereas for AB loops the error is significantly smaller across all scenarios. However, during the initial milliseconds of the fault (the crucial phasor calculation window), the impedance values are critically low, which can be further analyzed in the next scenario.

The case in Fig. \ref{fig:dist100} is important to analyze, as it represents an external fault at terminal 2, corresponding to 100\% of the line length. The behavior follows a similar pattern: the synchronous generation scenario shows the calculated impedance closest to the real value (dashed red line). Moreover, as previously stated, AG loops tend to be higher and outside the operating zone. However, when analyzing the AB loop, for a few milliseconds the apparent impedance values for all GFM, GFL, positive- or negative-sequence prioritization cases are positioned below the line reach, which could cause a misoperation if security thresholds do not block it. It is worth noting that prioritizing negative-sequence current results in a higher impedance.

On the other hand, the apparent impedance seen from the synchronous generation system is the only one that stays safely far from the 80\% line reach.

\subsection{Differential protection}

\begin{figure}[t]
    \centering
    \includegraphics[width=1\columnwidth]{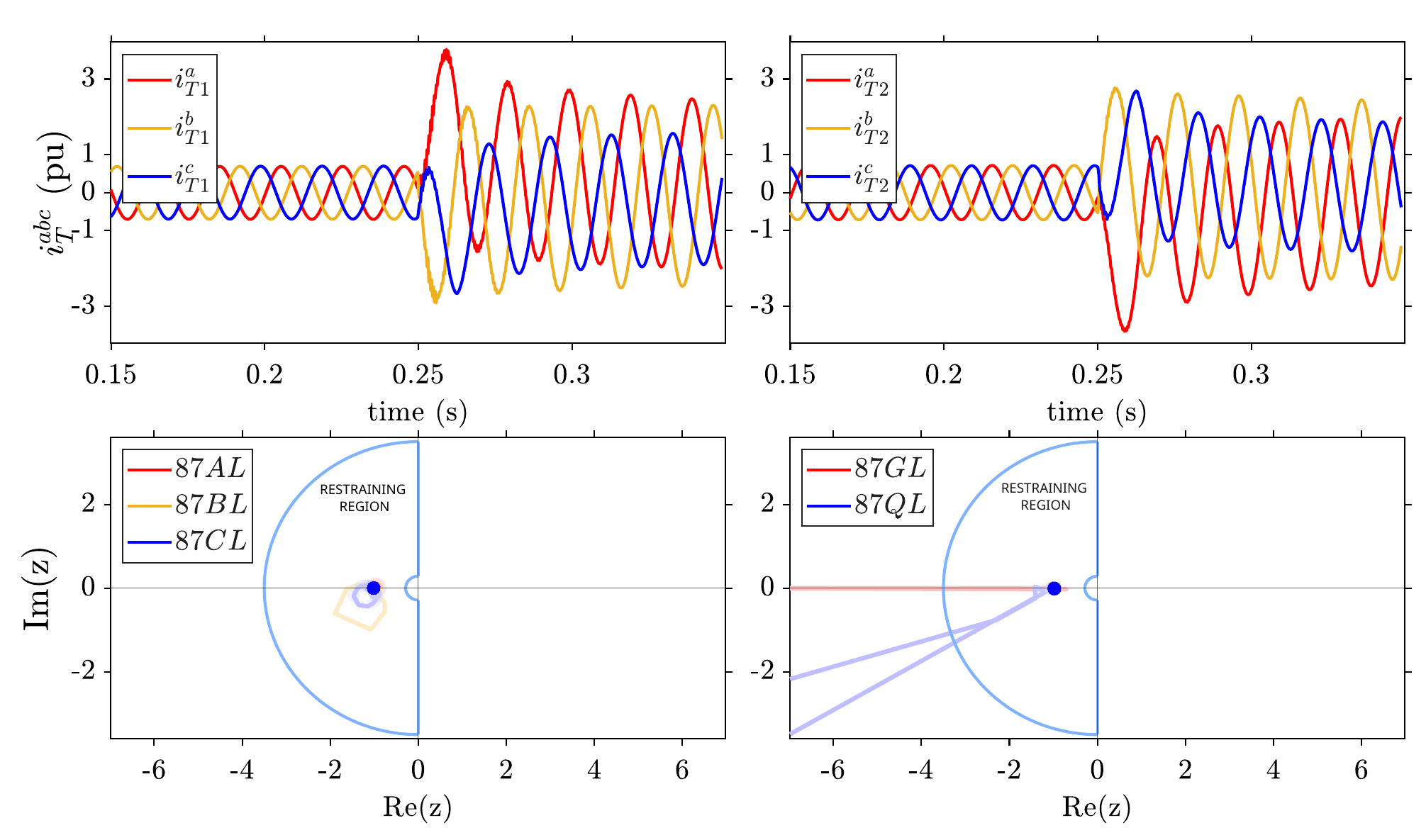}
    \caption{GFM with neg. prior. ABG fault at 0\% external}
    \label{fig:diff_negneg0_external}
\end{figure}

\begin{figure}[b]
    \centering
    \includegraphics[width=1\columnwidth]{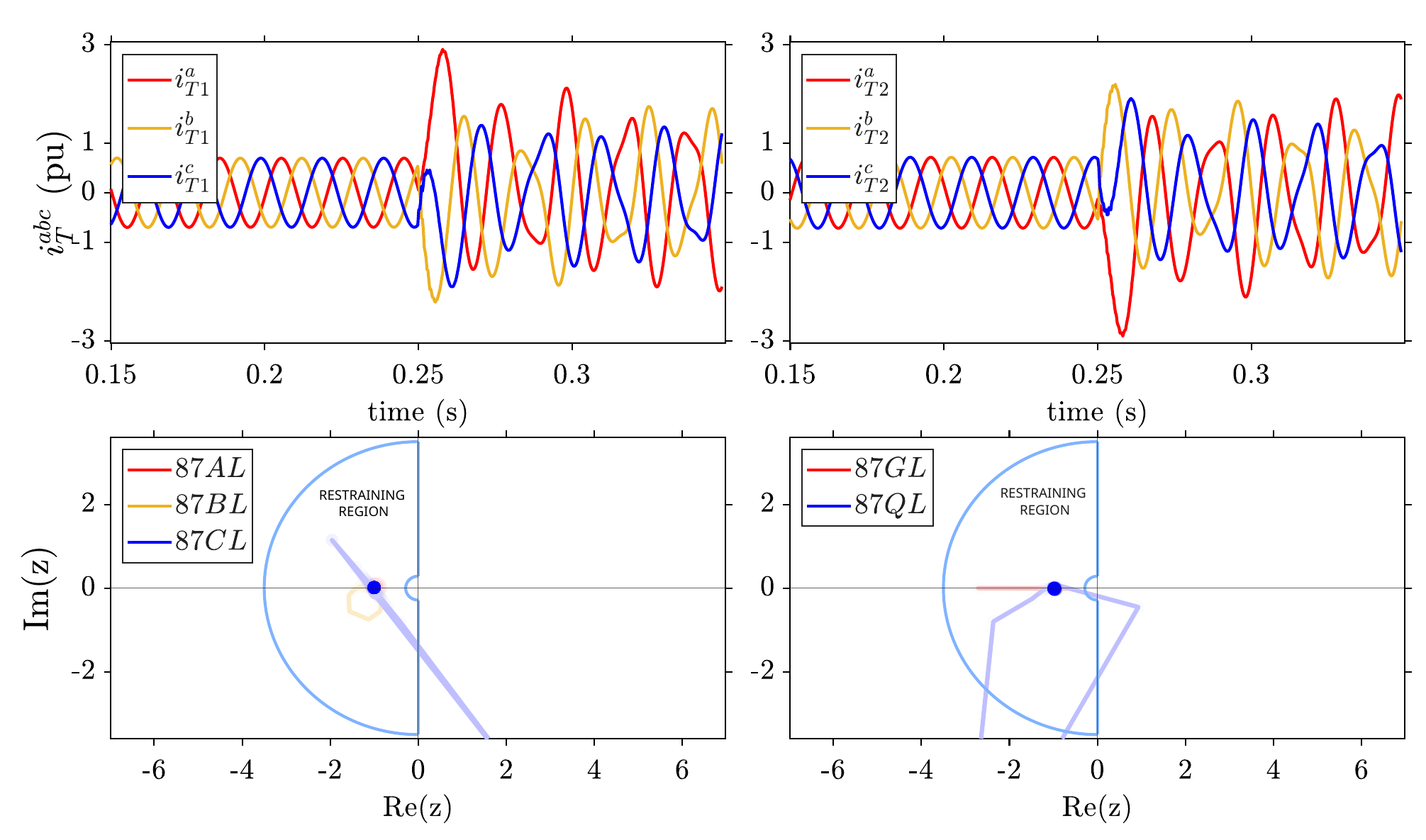}
    \caption{GFL with pos. prior. ABG fault at 0\% external}
    \label{fig:diff_pospos0gfl_external}
\end{figure}

Figure \ref{fig:diff_negneg0_external} shows the behavior of the mapping in the alpha plane for an external fault fed by the GFM scenario at terminal 1. The upper figures show the terminal currents. The bottom figures show the behavior of the differential quantities: the bottom-left plot shows the phase quantities $87LP$, and the bottom-right plot shows the ground $87GL$ and negative $87QL$ quantities.

It can be observed in Fig. \ref{fig:diff_negneg0_external} that the final differential quantities during the fault tend to remain inside the restraint zone (solid dots). However, for the zero- and negative-sequence quantities, some of the initial points during the fault move outside the zone, along the negative real axis.

\begin{figure}[t]
    \centering
    \includegraphics[width=1\columnwidth]{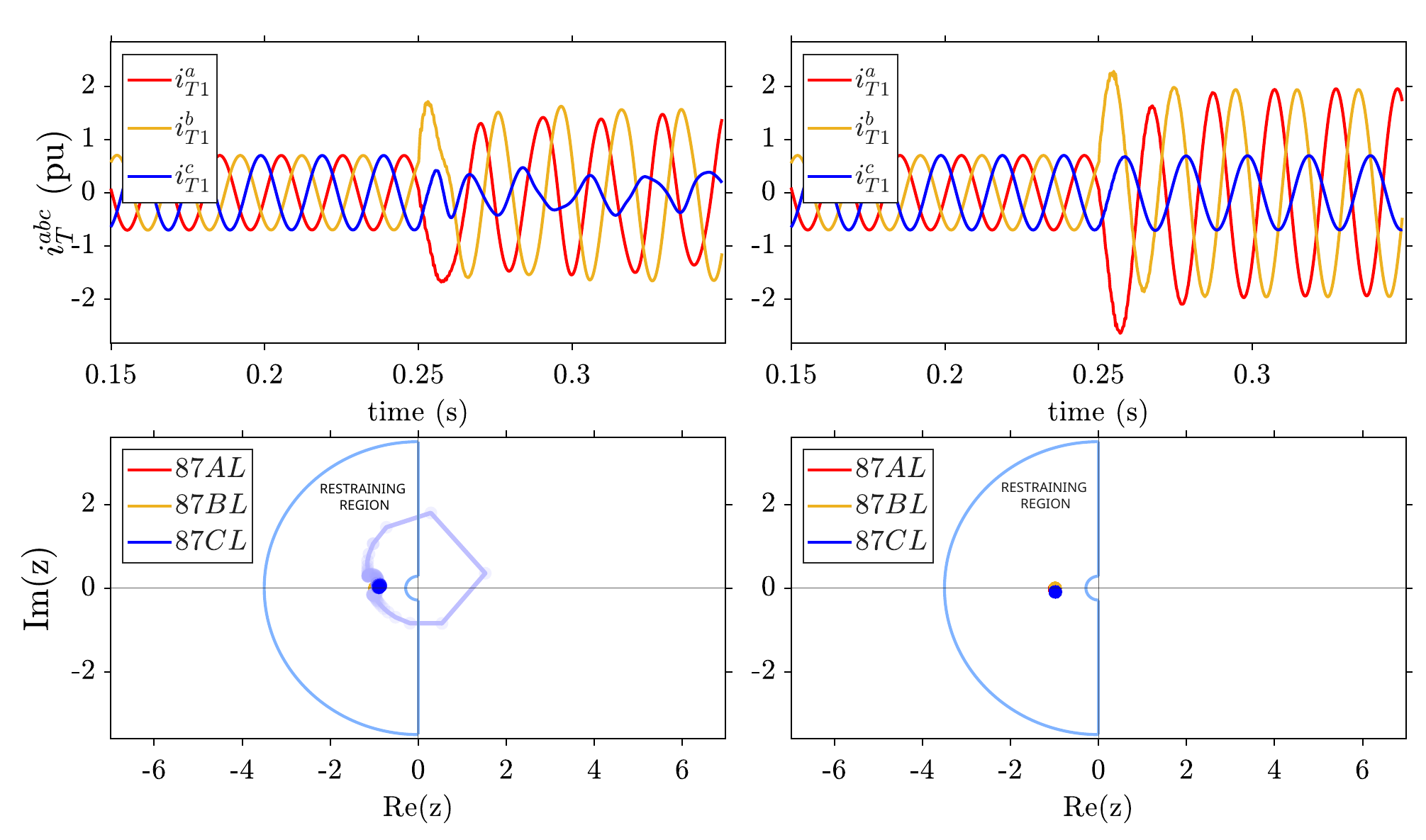}
    \caption{GFL with neg. prior. vs SG. ABG fault at 100\% external}
    \label{fig:diff_pospos100gfl_external}
\end{figure}

When comparing the GFM scenario with the GFL operation in Fig. \ref{fig:diff_pospos0gfl_external}, the electrical signals are highly distorted for the GFL operation, and more importantly, the zero- and negative-sequence differential quantities move into the right half of the alpha-plane for one sample. This is important, given that in these scenarios other security measures should prevent tripping. Moreover, not only the elements 87QL and 87GL show this behavior; in Fig. \ref{fig:diff_pospos100gfl_external}, an external fault at terminal 2 with $Z_F = 5 \ \Omega$ shows the GFL prioritizing negative-sequence current injection (left graphs) and is compared to the case with only synchronous generation (right graphs), this shows how the element 87CL (the healthy phase) moves into the right half of the alpha-plane for three samples before converging to the restraint region in the GFL scenario.

In general, for internal faults the differential algorithm correctly identifies faults with low and high fault resistances and for the different control schemes.

\section{Conclusions}

The continuous integration of inverter-based resources into transmission systems has affected their long-established operation, introducing new challenges in modern power systems, and this work provides evidence of that. In addition, new control schemes for integrating renewable resources through converters vary greatly, making the overall integration more complex. Moreover, the physical constraints of converters require current limiting and prioritization, which affect system performance.

This work has shown how line distance protection, one of the most widely used primary protection schemes today, is greatly affected by the different control strategies of converters, whether based on GFM or GFL control, or on different current-limiting and prioritization decisions. For faults near the limit of Zone 1, the apparent impedance can be larger than expected for LG loops, or exhibit transients that could result in overreaching for LLG loops. Also, prioritizing positive- or negative-sequence current injection could lead to different operation. Moreover, some important findings also arise for differential protection. Although generally considered a highly secure algorithm, for external faults at the busbar terminals and during the first instants of the fault, a few samples of the operating quantities may move into the right half of the alpha-plane, which is a trip region, requiring high security margins from other conditions to avoid an incorrect trip indication. These results support the idea that there is still a need to further explore protection algorithms, rather than assuming they are completely reliable when operating in modern power systems.

\vspace{-8 pt}
\section{Acknowledgments}
\vspace{-8 pt}
This publication and other research outcomes are supported by the predoctoral program AGAUR-FI ajuts (2025 FI-1 00374) Joan Oró, backed by the Secretariat of Universities and Research of the Department of Research and Universities of the Generalitat de Catalonia and European Social Plus Fund. Also, it has been supported by the HP2C-DT project (grant TED2021-130351B-C21). The work of Oriol Gomis-Bellmunt was supported by the Institució Catalana de Recerca i Estudis Avançats (ICREA).
\vspace{-8 pt}
\section{References}
\vspace{-8 pt}

\bibliography{references.bib}

\end{document}